# Learning differentially reorganizes brain activity and connectivity

Maxwell A. Bertolero[a], Azeez Adebimpe[b,c], Ankit N. Khambhati[a,d], Marcelo G. Mattar[e], Daniel Romer[c], Sharon L. Thompson-Schill[f], Danielle S. Bassett[a,c,g,h,i,j,k]

[a]Department of Bioengineering, School of Engineering & Applied Science, University of Pennsylvania, PA, 19104 USA
[b]Annenberg Public Policy Center, University of Pennsylvania, Philadelphia, PA, 19104 USA
[c]Department of Psychiatry, Perelman School of Medicine, University of Pennsylvania, Philadelphia, PA, 19104 USA
[d]Department of Neurological Surgery, University of California, San Francisco, CA 94122 USA
[e]Princeton Neuroscience Institute, Princeton University, Princeton, NJ, USA
[f]Department of Psychology, School of Arts & Sciences, University of Pennsylvania, PA, 19104 USA
[g]Department of Electrical & Systems Engineering, School of Engineering & Applied Science, University of Pennsylvania, PA, 19104 USA
[h]Department of Neurology, Perelman School of Medicine, University of Pennsylvania, Philadelphia, PA, 19104 USA
[i]Department of Physics & Astronomy, College of Arts & Sciences, University of Pennsylvania, Philadelphia, PA, 19104 USA
[j]Santa Fe Institute, Santa Fe, NM 87501 USA

[k]To whom correspondence should be addressed: dsb@seas.upenn.edu

Human learning is a complex process in which future behavior is altered via the reorganization of brain activity and connectivity. It remains unknown whether activity and connectivity differentially reorganize during learning, and, if so, how that differential reorganization tracks stages of learning across distinct brain areas. Here, we address this gap in knowledge by measuring brain activity and functional connectivity in a longitudinal fMRI experiment in which healthy adult human participants learn the values of novel objects over the course of four days. An increasing similarity in activity or functional connectivity across subjects during learning reflects reorganization toward a common functional architecture. We assessed the presence of reorganization in activity and connectivity both during value learning and during the resting-state, allowing us to differentiate common elicited processes from intrinsic processes. We found a complex and dynamic reorganization of brain connectivity and activity—as a function of time, space, and performance—that occurs while subjects learn. Spatially localized brain activity reorganizes across the brain to a common functional architecture early in learning, and this reorganization tracks early learning performance. In contrast, spatially distributed connectivity reorganizes across the brain to a common functional architecture as training progresses, and this reorganization tracks later learning performance. Particularly good performance is associated with a sticky connectivity, that persists into the resting state. Broadly, our work uncovers distinct principles of reorganization in activity and connectivity at different phases of value learning, which inform the ongoing study of learning processes more generally.

Although each human brain is unique, all brains share a common form. Functional connectivity, which measures the statistical similarity between the temporal activity of two brain regions, can vary appreciably across individuals, so much so that it serves as an individual's fingerprint[1]. Moreover, recent work has demonstrated that functional connectivity can predict task performance[2], including the capacity for skill learning[3–5]. Similarly, a single cognitively demanding task can elicit quite different patterns and magnitudes of activity in different individuals[6]. Finally, individual differences in functional connectivity during rest are related to individual differences in brain activity during task performance[7]. Despite these notable instances of variance in brain activity and connectivity, there also exist several invariable subject-general patterns; macro-organizational principles of functional connectivity appear to be conserved throughout healthy normative populations[8,9], and there are reliable group level task-induced organizations of brain activity[10] and functional connectivity[9].

Similarly, learning is a dynamic process with both similarities and differences across individuals; everyone learns differently, but much of the mechanics of learning follow a common form. While individual differences in brain and behavior certainly exist, learning reorganizes neural processes to be markedly similar across individuals for various reasons: stimuli present during learning, the mechanics of the particular skill that is learned, the information that one learns, and learning strategies, if similar across subjects. Moreover, conserved physiological and anatomical organization across subjects can foster that reorganization of brain activity and connectivity to a common functional architecture[11]. Finally, shared genetic and environmental factors present in the early stages of development can give rise to common wiring patterns[12], which can in turn lead to conserved spatiotemporal responses in neuronal ensembles[13].

While we know that individuals have unique brains that still share a common form, and that task engagement can reorganize brain activity and connectivity to a common functional architecture, it is not known how that reorganization varies across the brain or tracks distinct stages of learning. Further, it is unknown if activity and connectivity differentially reorganize during learning. Notably, while resting-state connectivity has been shown to shape task activity[14], it is possible that reorganizations of activity and connectivity emerge at distinct brain regions and distinct time points during learning, as functional connectivity during a task can be distinct from activity, with the activity reorganization being more directly driven by stimulus encoding[15]. To better understand the brain's reorganization to a common functional architecture that accompanies learning, we must determine the temporal evolution and spatial extent of similarities and differences in activity and connectivity during learning. Increased similarities during learning reflect the extent to which activity or connectivity have been reorganized by learning processes to a common functional architecture that may support learning. Accordingly, here we use inter-subject correlation and inter-subject functional connectivity to assess the extent to which each region's activity and connectivity exhibit individual similarities (high inter-subject correlation or inter-subject functional connectivity) or differences (low inter-subject correlation or inter-subject functional connectivity) across subjects as they learn the value of novel objects over the course of four consecutive days (Figure 1). We contrast our findings with those obtained from the resting state to differentiate between common processes elicited by learning and intrinsic processes occurring endogenously.

**Measuring reorganization in activity and connectivity during learning.** To probe the extent of network reorganization during learning, we investigated inter-subject correlation and inter-subject functional connectivity (Figure 2) estimated from functional MRI (fMRI) data acquired in 20 healthy subjects (9 females; ages 19–53 years; mean age = 26.7 years) over four consecutive days (Figure 1d), during which learning significantly increased from day one to day four (Figure 1e). Each scanning session contained data collected while the subject rested and while the subject engaged in a value-learning task (Figure 1a-c). From each session, we extracted the BOLD time series of 400 cortical (Schaefer-Yeo)[16] and 14 subcortical regions (Harvard-Oxford)[17,18]. We measured inter-subject correlation by calculating the correlation between regional BOLD signal time series for each pair of 20 subjects. This procedure resulted in a 20x20 correlation matrix for each brain region, the mean of which is the region's inter-subject correlation.

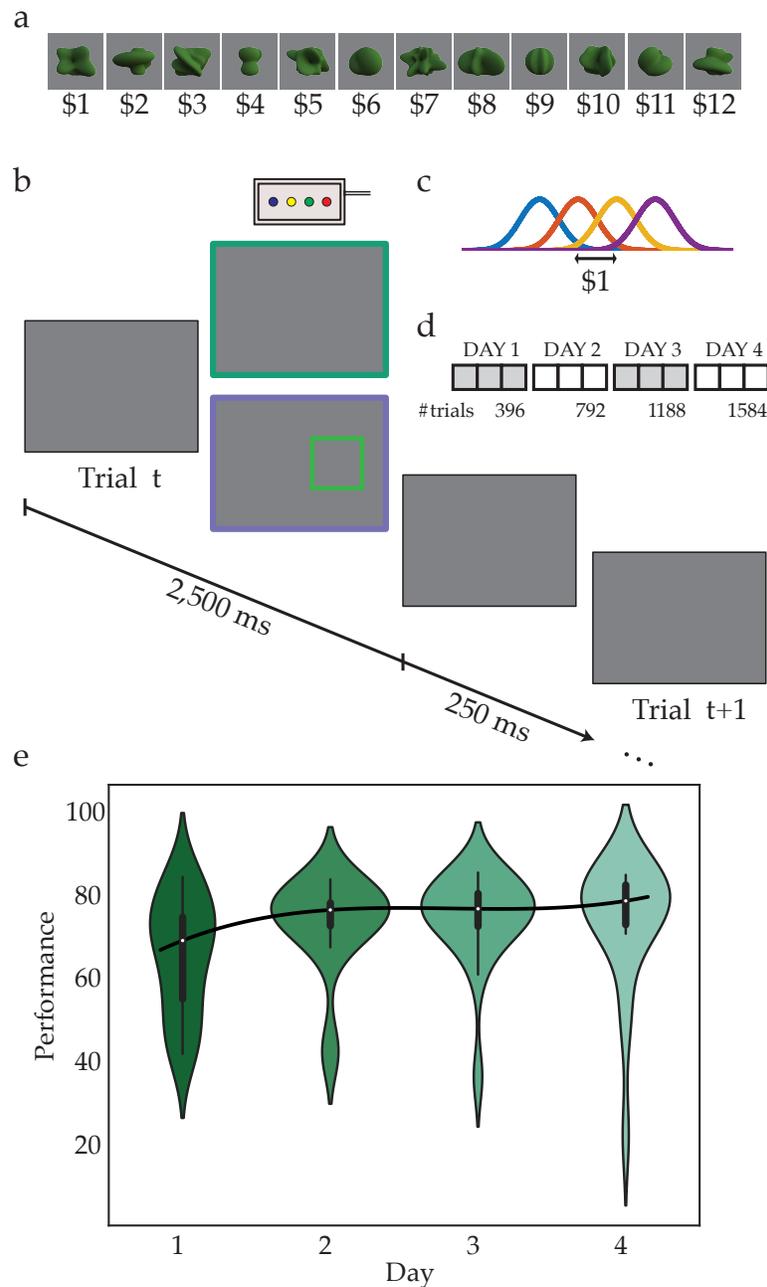

**Figure 1 | Learning task and performance**. **a**, Stimulus set and corresponding values. Twelve abstract shapes were computer generated, and an integer value between $1 and $12 was assigned to each. This value remained constant over the four days of training. **b**, Participants were presented with two shapes side-by-side on the screen and asked to choose the shape with the higher monetary value. Once a selection was made, either the value of the shape selected (absolute feedback) or the correctness of the selection (relative feedback) was provided as feedback. Each trial lasted 2,750 ms (250 ms interstimulus interval). **c**, On each trial, the empirical value of each shape was drawn from a Gaussian distribution with fixed mean (i.e., the true value), as described in panel (**a**), and standard deviation of $0.50. **d**, The experiment was conducted over four consecutive days, with three experimental scans (396 trials) on each day, for a total of 1,584 trials. **e**, The distribution of performance values for each subject on each day is shown. Median subject performance increased as a function of day, and there was a significant increase in learning performance from day one to day four (relative $t$-test=2.24, $p$=0.03, $df$=38). Figure adapted with permission from a previous study[19].

Next, we constructed a functional connectivity matrix for each subject and each scan, where each $ij^{th}$ element in the 414x414 matrix indicated the wavelet coherence between the time series of region $i$ and the time series of region $j$. We then measured inter-subject functional connectivity for a given subject and region $i$ by computing the Pearson correlation between row $i$ in that subject's functional connectivity matrix and the average of row $i$ in all of the other subjects' functional connectivity matrices. We performed this calculation for every region, resulting in an inter-subject functional connectivity array of length 414 for each subject (see Materials and Methods). Thus, in sum, we have a measure of each region's activity and connectivity reorganization in each subject on each day of learning.

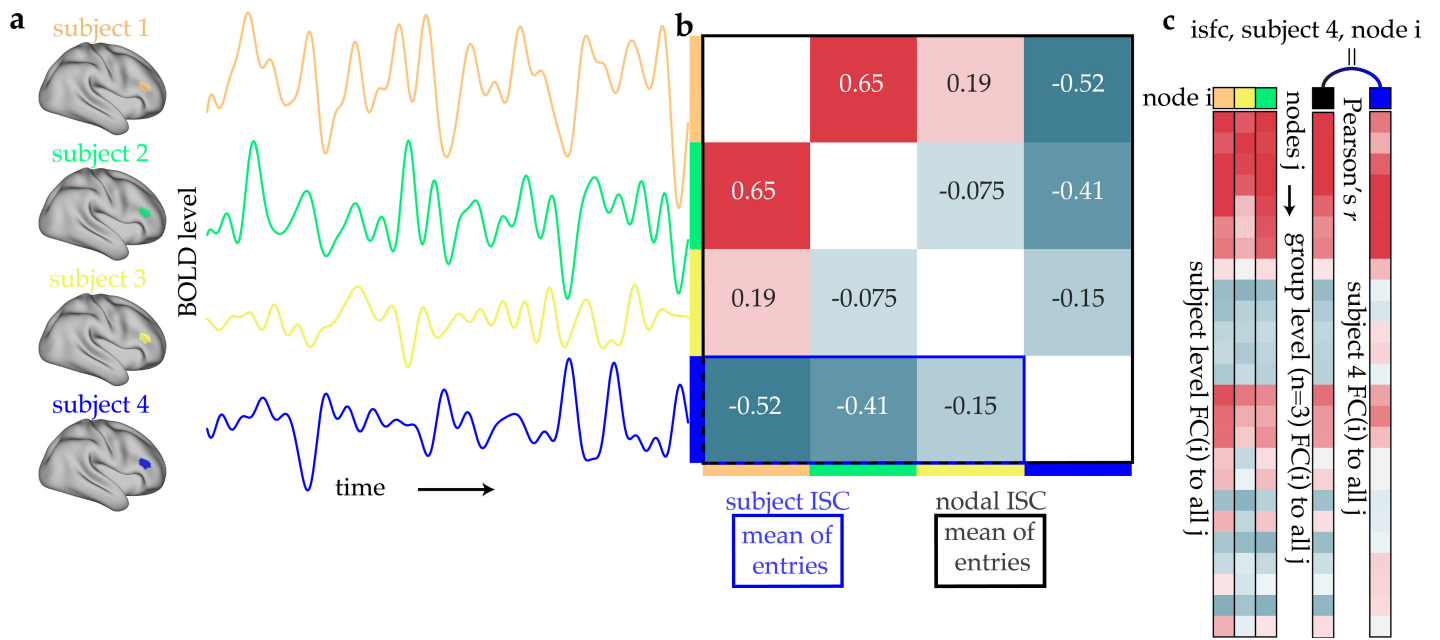

Figure 2 | **Measuring inter-subject correlation and inter-subject functional connectivity**. **a** To measure inter-subject correlation (ISC) and inter-subject functional connectivity (ISFC), the mean time series of a single region (i.e., network node; here, node *i*) is extracted in each subject. Here we provide an illustrative depiction with four subjects for node *i*. **b** The correlation between all pairs of subjects' time series is calculated. The mean of all entries (black box) is the region's inter-subject correlation. The mean of a subject's row (blue box) is the subject's inter-subject correlation. **c**, To measure inter-subject functional connectivity for a single subject (subject #4) and region *i*, the functional connectivity between region *i* and all other regions *j* is calculated for each subject. Next, the mean connectivity between region *i* and all other regions *j* is calculated without subject #4 (black box). The inter-subject functional connectivity for subject #4 and region *i*, then, is the Pearson correlation between the mean connectivity between region *i* and all other regions *j* (without subject #4; black box) and the connectivity between region *i* and all other regions *j* in subject #4 (blue box).

We began by determining which brain regions exhibited inter-subject correlation values that were significantly greater than expected across all task sessions (one-sample *t*-test, $p < 0.05$, FWER corrected for multiple comparisons with 10000 random permutations[20], $df=7866$). We observed significant inter-subject correlation broadly across the brain (Figure 3a shows significant regions). Next, we determined which brain regions exhibited inter-subject functional connectivity values that were significantly greater than expected across all task sessions (one-sample *t*-test, $p < 0.05$, FWER corrected for multiple comparisons with 10000 random permutations[20], $df=7866$). We also observed significant inter-subject functional connectivity broadly across the brain (Figure 3b shows significant regions). The spatial differences in inter-subject correlation and inter-subject functional connectivity across the brain were investigated further by *z*-scoring the values of each and subtracting them from one another (Figure 3c,d). These results demonstrate that many brain regions exhibit a reorganization of connectivity and activity during the learning of value, but there are spatial differences in where reorganizations in activity versus connectivity appear.

Next, we asked whether regions of relatively high inter-subject correlation versus inter-subject functional connectivity were also previously associated with the encoding and representation of value during learning. To determine which regions were associated with value, we used the "value" term association map in

Neurosynth[21], which reflects how often each voxel is reported in studies that contain the word "value" in the abstract ($n = 470$) in comparison to all other studies. Because the Neurosynth map is produced at the voxel level but our inter-subject measurements are produced at the region level, we coarse-grained the Neurosynth map by calculating the across-voxel mean for each brain region. The regions ($n=10$) most associated with value learning have higher inter-subject functional connectivity than inter-subject correlation on average during learning ($t=2.65$, $p=0.02$, $df=18$; non-parametric permutation based one-sample $t=76.9$, $-\log10(p)>100$ for $n=1000$ permutations; Figure 3c,d). Thus, for each of the value-associated regions, even though it is reliably activated during the learning of value, that region's connectivity to the rest of the brain is more similar across subjects than its time series of activity across subjects.

While differences existed between the brain regions displaying high inter-subject correlation and the brain regions displaying high inter-subject functional connectivity, we did observe a positive correlation between regions' inter-subject functional connectivity and inter-subject correlation ($r=0.42$, $-log10(p)=19$, $df=412$). Moreover, within each subject, we observed a positive correlation between regions' inter-subject functional connectivity and inter-subject correlation ($r$(mean across subjects)$=0.28$, $p$(mean across subjects)$=0.026$, $df=412$). However, the extent to which an individual displayed high inter-subject functional connectivity was not correlated with the extent to which that individual displayed high inter-subject correlation ($r=0.203$, $p>0.05$, $df=78$). Thus, whereas the extent to which each subject's activity and connectivity reorganize globally are unrelated, the extent to which each region's activity and connectivity reorganize are similar, both within and across subjects.

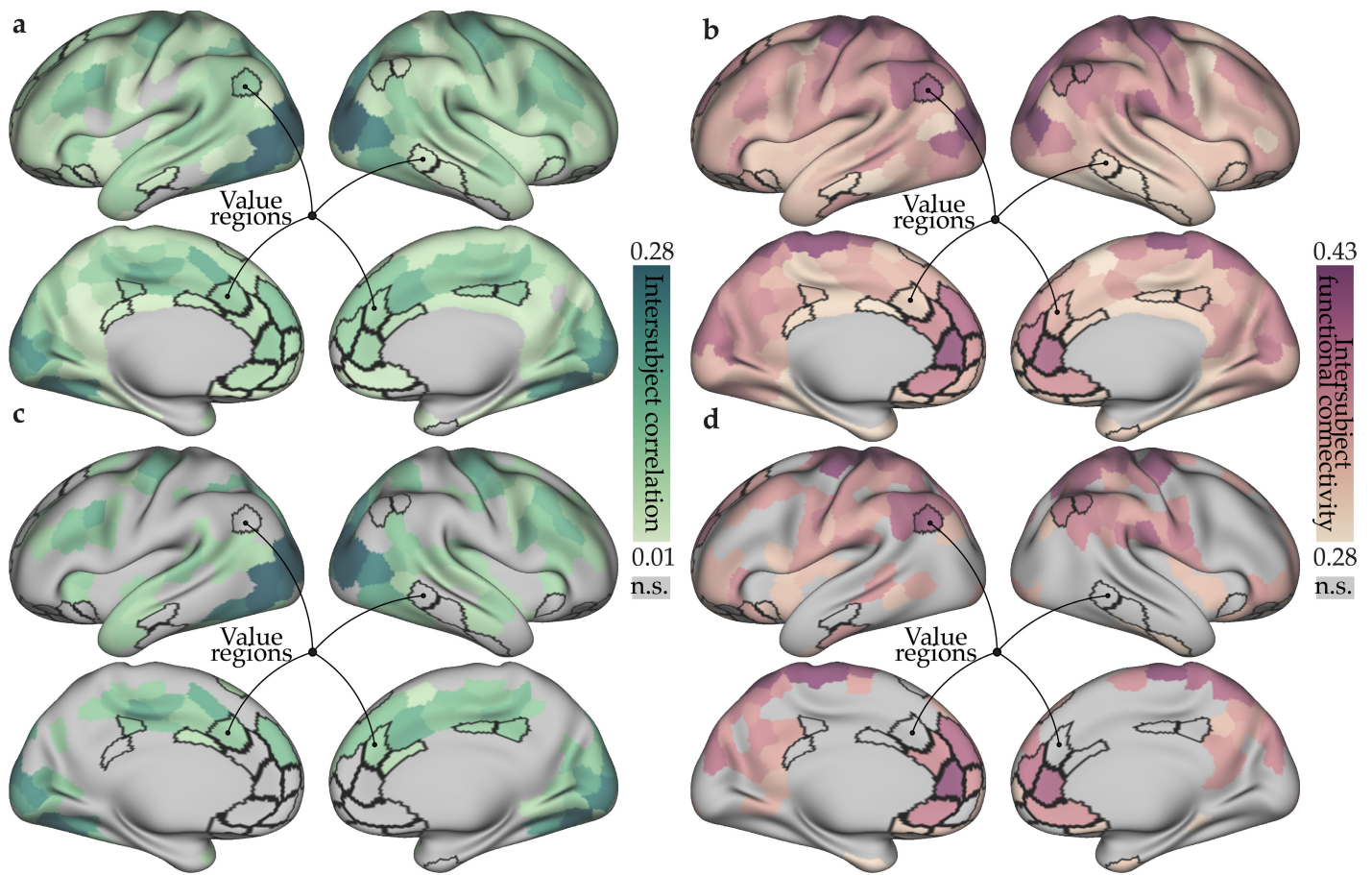

Figure 3 | **Activity and connectivity reorganization during learning**. **a,** We observed significant inter-subject correlation, reflecting activity reorganization, across most of the brain during the performance of the value learning task. Regions that did not pass multiple comparisons correction are shown in grey. Regions associated with value in a Neurosynth meta-analysis are outlined in black. **b,** We observed significant inter-subject functional connectivity, reflecting connectivity reorganization, across most of the brain. **c,d,** To visualize the spatial differences in activity and connectivity reorganizations,

the values were *z*-scored separately and then subtracted from one another. The regions most associated with value learning have higher connectivity than activity reorganizations (*n*=9 regions; *t*=2.65, *p*=0.02, df=18).

**Early learning reorganizes activity, whereas later learning reorganizes connectivity.** Next, we asked if similarities in activity and functional connectivity across subjects, reflecting reorganization toward a common architecture, could distinguish between different phases of value learning. We reasoned that activity reorganization, reflected by higher inter-subject correlation, is likely to be high early in learning when value is being encoded and stimulus processing is most prominent[15]. In contrast, we reasoned that reorganization in connectivity, reflected by higher inter-subject functional connectivity, is likely to occur later in learning as more widespread cortical networks engage more stereotypically in the computations that support learning and memory consolidation of the object values[4,22–24]. To test these hypotheses, we first examined the temporal dynamics of inter-subject correlation during value learning (Figure 4a). We found that the average inter-subject correlation over all brain regions increased from the first day to the second day (*t* = 10.32, −*log*10(*p*) = 25, *df* = 8278), and then subsequently decreased through the fourth day (*t* = 13.99, −*log*10(*p*) = 44, *df* = 8278; Figure 4a). In a complementary analysis, we examined the temporal dynamics of inter-subject functional connectivity, which we found increased from day one to day two (*t* = 3.01, *p* = 0.002, *df* = 8278), from day two to day three (*t* = 2.91, *p* = 0.003, *df* = 8278), and from day three to day four (although not by a significant amount; Figure 4b).

Next, we explicitly tested for potential interactions between reorganization in activity and connectivity over time. We uncovered evidence for a divergence in reorganization of activity and connectivity during the later stages of learning; for these comparisons we *z*-scored the inter-subject functional connectivity and inter-subject correlation values, so that each region has a mean of zero across days. Specifically, on day two, we observed that inter-subject functional connectivity values were significantly lower than inter-subject correlation values (*t* = 2.2, *p*=0.039, *df* = 18; Figure 4c), and, on day four, inter-subject functional connectivity values were significantly greater than inter-subject correlation values (*t* = 3.2, *p*=0.005, *df* = 18; Figure 4c). These results demonstrate that reorganization in activity peaks early in learning, while reorganization in connectivity slowly increases throughout learning.

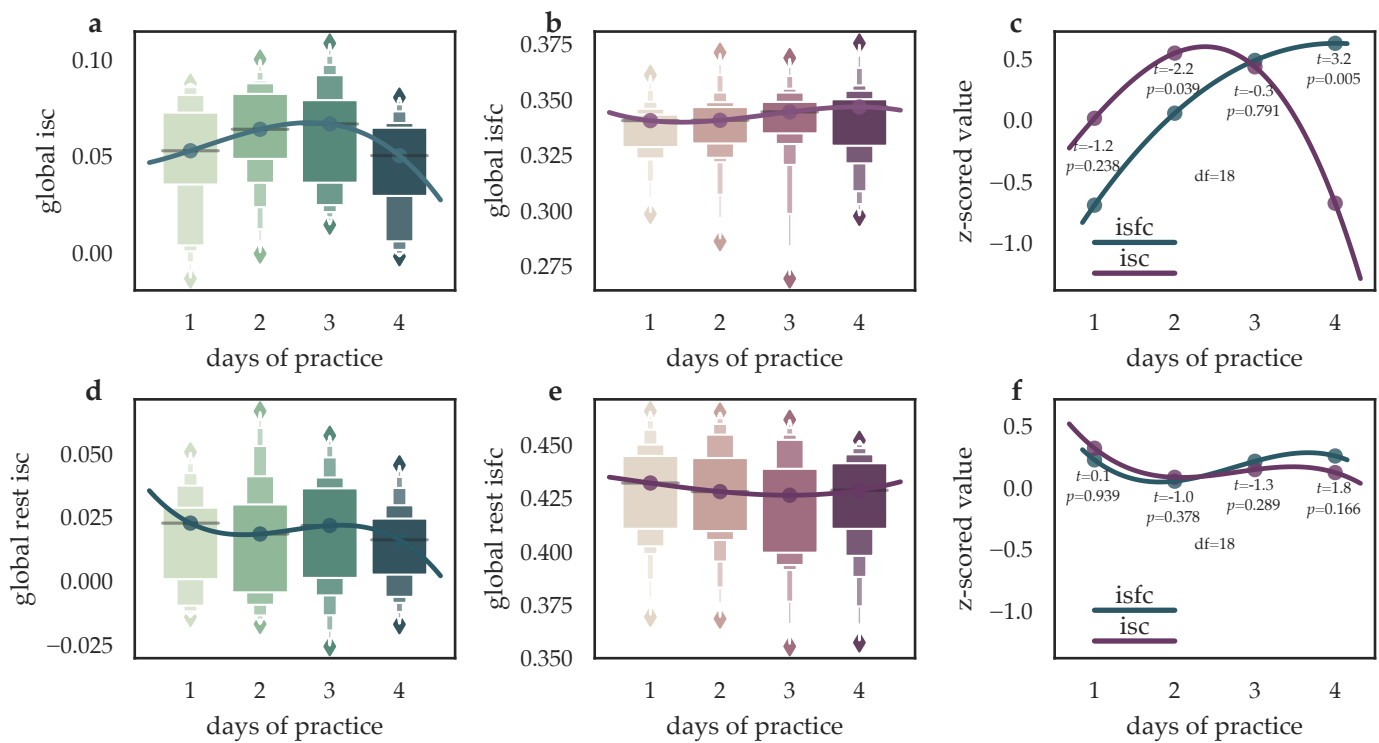

Figure 4 | **Global reorganization of activity and connectivity during learning. a**, We observed that inter-subject correlation, reflecting activity reorganization, increased from day one to day two of task practice, and then decreased in day

three and day four. **b**, We observed that the inter-subject functional connectivity, reflecting connectivity reorganization, increased from day one to day two of task practice, as well as from day two to day three, thereafter remaining relatively steady. **c**, Median values for *z*-scored activity and connectivity reorganization are shown. Whereas activity reorganization peaks on day two, connectivity reorganization peaks on days three and four. **d–f**, Same as in panels **a–c**, but in this case the activity and connectivity reorganizations were measured during the resting state rather than during task performance. Note that the temporal dynamics observed during learning were not observed during the resting state.

It is important to ask whether these same temporal trends in inter-subject correlation and inter-subject functional connectivity would have occurred outside of a learning context. To address this question, we constructed a null model by computing inter-subject correlation for the resting-state condition, during which fMRI data was also acquired in each of the four days of the experiment. In contrast to the inter-subject correlation during value learning, we did not observe a significant increase from day one to day two, but we did observe a decrease, albeit much smaller, from day three to day four (Figure 4d, $t=2.42$, $p=0.01$, $df=8278$). Moreover, in contrast to inter-subject functional connectivity during value learning, we observed that the inter-subject functional connectivity during rest actually decreased from days one to two and three ($t=6.92$, $-\log10(p)=12$, $df=8278$) and then increased on day four ($t=2.09$, $p=0.003$, $df=8278$; Figure 4e). Critically, in contrast to the value learning condition, *z*-scored inter-subject correlation and inter-subject functional connectivity values exhibited conserved temporal dynamics during rest, with no significant differences on any day (Figure 4f). Thus, our results regarding the temporal evolution of inter-subject correlation and inter-subject functional connectivity are driven by learning, not time, as they are only observed during the learning condition, and not during the resting-state condition.

In sum, from day one to day two of learning, both connectivity and activity reorganize to a common functional architecture that putatively supports learning, with activity reorganizing to a greater extent. Later in learning (days three and four), connectivity continues to reorganize, while activity does not. These dynamics of higher activity reorganization on days one and two, and higher connectivity reorganization on day four, suggest a potential driver-follower mechanism whereby reorganization in activity precedes reorganization in functional connectivity. The increase in connectivity reorganization later in learning may in part reflect the consolidation of memories for the values experienced during learning.

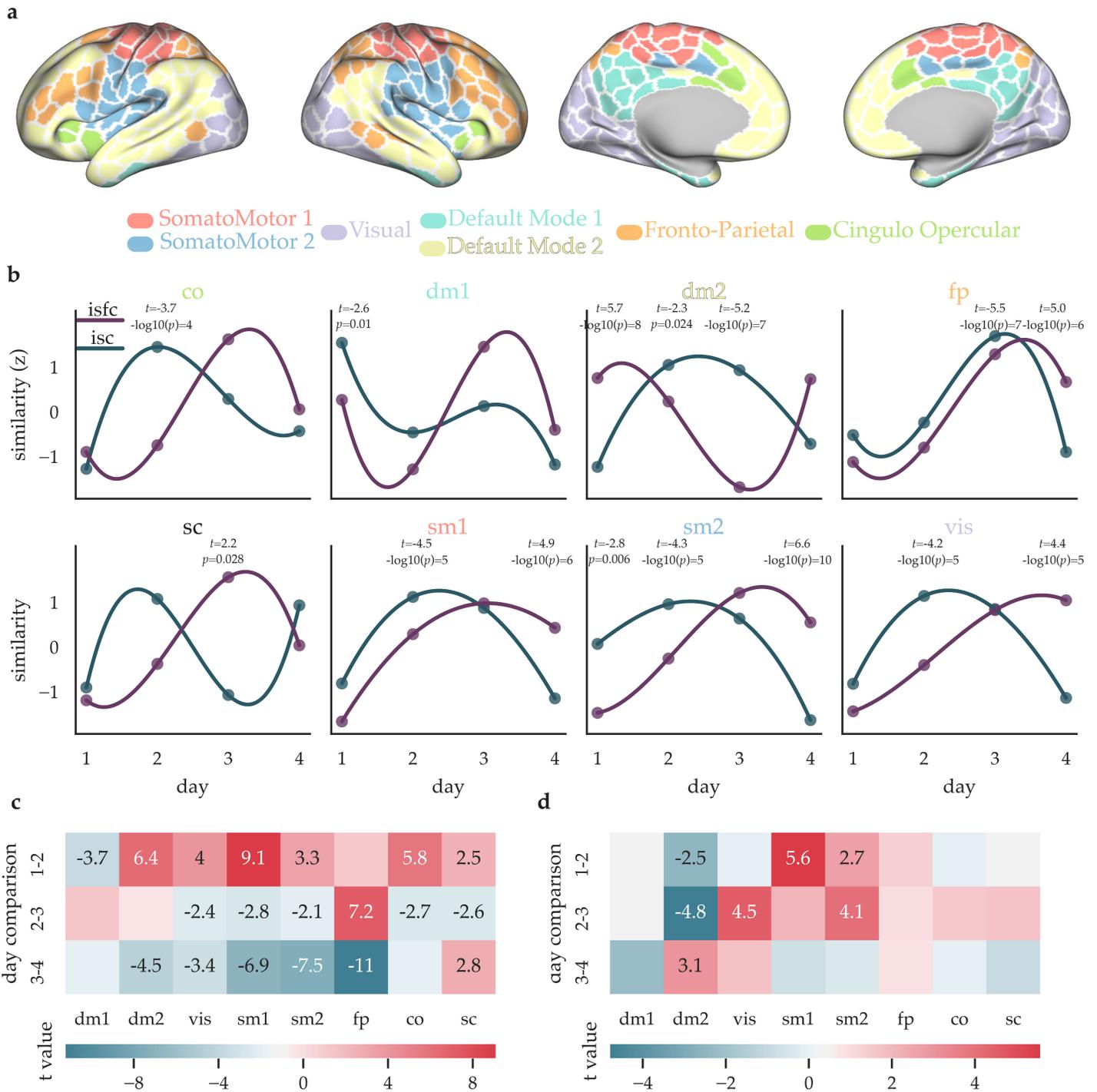

**Figure 5 | Community level reorganization of activity and connectivity during learning**. **a**, To examine meso-scale structure in the whole-brain patterns of activity and connectivity reorganizations, we detected communities in the data acquired during the performance of the value learning task. **b** For each functional community, the median activity and connectivity reorganization across nodes in that community is shown for each day of learning. **c**, We compared activity reorganization values between each day of learning in each brain community. Significant *t*-values from a relative *t*-test, FDR Benjamini/Hochberg corrected, are shown for each comparison. **d**, As in panel **c**, but for connectivity reorganizations.

**Learning reorganizes activity globally and reorganizes connectivity locally.** If reorganization of activity and connectivity occurs at different brain regions and during different phases of value learning, it is natural to ask which individual brain systems exhibit this phenomenon. To address this question, we first partitioned the set of 414 regions into functional modules by applying the GenLouvain community detection algorithm[25] to the functional connectivity matrices (see Methods). The resultant data-driven partition was comprised of eight communities, which we interpret as putative functional communities, in line with prior work[16,26] (Figure 5a). We found that the inter-subject correlation was significantly higher in the visual and somatomotor communities than in other communities, and was significantly lower in the default mode and subcortical communities than in other communities ($2.88 < t > 39.93$, $p < 0.05$ after FDR correction across $n=8$ tests, df=33118). We also found that the inter-subject functional connectivity was significantly higher in the visual, somatomotor, and fronto-parietal communities than in other communities ($7.02 > t < 38.63$, $p < 0.05$ after FDR correction for $n=8$ tests, df=33118). This pattern of results demonstrates that functional connectivity and activity reorganize most in somatomotor communities, with functional connectivity additionally reorganizing in the fronto-parietal communities, perhaps because of the need for the fronto-parietal network to coordinate and integrate connectivity across somatomotor regions[2,4,27–29]. Generally, this finding is consistent with the notion that functional connectivity is distinct from activity, with the latter being more driven by stimulus encoding[15].

Although the average values of inter-subject correlation and inter-subject functional connectivity on each day of learning are an important place to start, when studying a complex process such as learning, it is also important to evaluate the time evolution of reorganization at the community level. Accordingly, we next assessed the degree to which the time-dependent variation in reorganization might differ across the eight brain communities. For every community, the inter-subject correlation increased from day one to day two, except for the first default mode community (Figure 5c). In contrast, the global increase in inter-subject functional connectivity from day one to day two is primarily driven by the two somatomotor communities and the fronto-parietal community (Figure 5d). More widespread increases in inter-subject functional connectivity occurred between days two and three, across all communities except the default mode communities. On the final day of learning, the default mode community was the only community to display significantly increased inter-subject functional connectivity. Finally, the only community to increase in inter-subject functional connectivity throughout learning, which was the temporal evolution observed at the global scale, was the fronto-parietal community.

Taken together, this pattern of findings implies that the temporal evolution of reorganization in activity occurs across all communities in the brain. However, for connectivity, reorganization during learning appears to be mostly driven by fronto-parietal and somatomotor communities, with default mode regions contributing in late learning. The result is particularly interesting in light of the need for the fronto-parietal network to coordinate and integrate connectivity across unimodal regions[2,4,27–29]. In contrast to connectivity, highly similar activity early in learning may reflect stimulus encoding that is highly similar across the brains of all subjects[15]. The final increase in reorganization of connectivity in the default mode community might reflect memory consolidation of the objects' values; in fact, this default mode community was largely comprised of the temporal lobe and regions previously associated with value.

**Reorganization of brain connectivity becomes spatially distributed during learning.** We next sought to examine the temporal evolution of reorganization at the level of single brain regions. We began by visualizing each region's activity and connectivity reorganizations on each day of learning (Figure 6a,b). The changes in regional activity reorganization largely replicate the pattern of findings at the community and whole-brain scales. Specifically, activity reorganization consistently peaks early in learning across brain regions. Notably, this feature is particularly evident in regions associated with the learning of value, such as orbital frontal cortex[15] on day two (Figure 6a). In contrast, connectivity reorganization displays a pattern of regional variation that offers further insight not accessible at the community and whole-brain scales. Specifically, the regional connectivity reorganizations display a spatial distribution that evolves throughout learning: early in learning, patches of high connectivity reorganizations appear locally constrained, being evident predominately in the fronto-parietal and default mode communities, while later in learning, regions with high connectivity reorganization appear globally spread across the brain. We statistically quantified this change in the nature of the connectivity reorganizations from local to global by calculating the Moran's $I$ statistic, a measure of the spatial autocorrelation in data based simultaneously on feature locations and feature values. We found that $(-1)$*Moran's $I$ significantly increased during learning for connectivity reorganizations but not for activity reorganizations ($t = 2.61$, $p = 0.012$, $df = 38$; Figure 6a-e), and, critically, these effects were not evident during the resting-state ($t=-0.6$, $p=0.523$, df=1654). Moreover, we found a similar function to global temporal dynamics in activity and connectivity reorganizations during learning; here, activity reorganization was more spatially spread across the brain than connectivity reorganization on the first day of learning, and the inverse was true on the last day of learning (Figure 6a-e). Critically, these effects were not observed during the resting state (*p*>0.1, df=1654). Finally, both activity and connectivity reorganizations were significantly more spatially distributed during learning than during rest (inter-subject correlation, 2.21 < t > 3.93, p<0.033; inter-subject functional connectivity, 4.9 < t > 6.76, -log10(p)<5, df=38). This pattern of findings demonstrates that value learning induces a more widespread reorganization in functional connectivity than in activity.

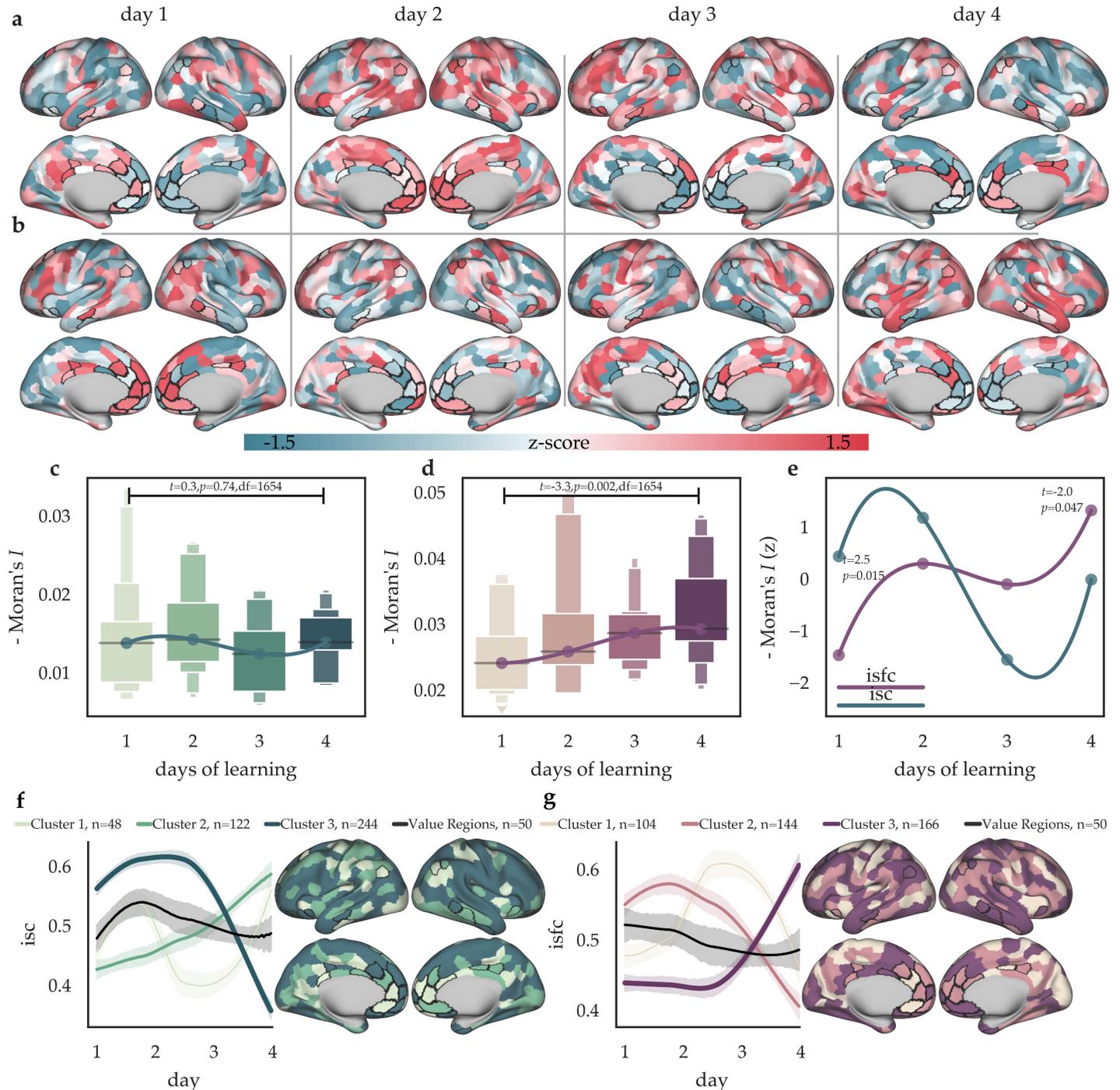

Figure 6 | **Regional reorganization of activity and connectivity during learning**. **a,b**, For each day, we show every region's inter-subject correlation (**a**) and inter-subject functional connectivity (**b**), reflecting activity and connectivity reorganization, respectively, where each node's inter-subject correlation or inter-subject functional connectivity sums to zero across days (*z*-scored). The overall decrease in activity reorganization can be readily observed at the regional level, with the lowest levels occurring on the last day of learning. Moreover, for connectivity reorganization, throughout learning, the spatial clustering of regions with high connectivity reorganization decays, as high connectivity reorganization becomes more distributed across the brain. **c,d,e**, We quantified this change in the spatial statistics using Moran's *I*, which measures

the extent of spatial autocorrelation from clustered to disperse. A high inverse Moran's *I* suggests high dispersion and low clustering. The - Moran's *I* significantly decreases during learning for connectivity reorganization (**d**) but not for activity reorganization (**c**). **f,g** To determine whether groups of regions display distinct temporal dynamics of reorganization during value learning, we used *k*-means clustering to partition brain regions into 3 groups that each displayed distinguishable activity and connectivity reorganization dynamics. The mean time courses of each cluster are shown in shades of green for activity reorganization and in shades of purple for connectivity reorganization, with shaded areas representing 95 percent confidence intervals. Additionally, the mean time courses for activity and connectivity reorganization in regions that have previously been associated with the learning of value in prior literature are shown in black. Activity reorganization, both in the largest cluster and in value regions, peaks early during learning. While the largest cluster for connectivity reorganizations peaks late in learning, the two other clusters peak earlier in learning, and value regions slowly decrease in connectivity reorganizations during learning.

Finally, we sought to determine whether reorganization of activity or connectivity manifested distinct modes of temporal dynamics. To address this question, we first generated a correlation matrix, where the *ij*-th entry represents the correlation between node *i*'s reorganization values and node *j*'s reorganization values throughout learning, using PCHIP 1-d monotonic cubic interpolation to generate 100 time points. We used *k*-means clustering (sklearn.cluster.**KMeans**) and identified $k = 3$ sets of brain regions, where each set was composed of regions that exhibited similar time courses of change in reorganization of activity (or, separately, connectivity) during learning. In a distinct analysis, we also examined the time courses of regions that had previously been associated with the learning of value in prior literature. For activity reorganization, the regions previously associated with the learning of value, and the regions located in the largest data-driven cluster ($n = 238$) identified by *k*-means, both exhibited a peak in reorganization on day two of learning and a subsequent decrease thereafter (Figure 6c). The second data-driven cluster composed of $n = 108$ regions displayed an increase in reorganization on day one, a decrease in reorganization on day three, and then an increase in reorganization on day four. The third data-driven cluster composed of $n = 68$ regions displayed an increase in reorganization throughout learning. For connectivity reorganizations, the $k = 3$ data-driven clusters reflect distinct dynamics. Two clusters of regions ($n=98$, $n=132$) display an early peak in connectivity reorganizations followed by a decrease in connectivity reorganizations towards the end of the fourth day of training. In contrast, the largest cluster composed of $n = 238$ regions displayed a constant level of connectivity reorganization during early learning and a peak during late learning. Finally, the regions that had previously been associated with the learning of value in prior literature display a slow but steady decrease in connectivity reorganizations.

In sum, reorganizations in activity are spread across the brain early in learning, while reorganizations in connectivity are spread across the brain throughout learning. Moreover, in most regions, and certainly in value learning regions, reorganization of activity peaks early in learning whereas reorganization of connectivity cascades, with different sets of regions reorganizing on each day of learning.

**Early learning performance is explained by activity reorganization, whereas later learning performance is explained by connectivity reorganization**. Lastly, we asked whether reorganization in activity and functional connectivity could explain interindividual differences in learning. For each region, we calculated the Pearson correlation coefficient, *r*, across subjects, between each region's activity reorganization (or connectivity reorganization) and task performance on that day. We refer to each *r* value as the learning coefficient for that region. Learning coefficients for activity and connectivity follow a similar temporal pattern as activity and connectivity reorganizations in general (Figure 4, Figure 5). Connectivity learning coefficients are significantly higher than activity learning coefficients on day one. Then, activity learning coefficients are marginally higher on days two and three. Finally, connectivity learning coefficients are higher on the last day of learning (Figure 7a-c). The higher connectivity learning coefficients on day one might reflect the existence of a connectivity pattern similar to the group-level that prepares one for learning. As learning progresses (days two and three), connectivity patterns are more divergent across subjects than activity patterns. Because activity patterns reflect stimulus encoding, among other processes, activity reorganization to the common functional architecture may be more useful than connectivity reorganization early. In contrast, by the last day of learning, connectivity

patterns similar to the group level are most related to learning, perhaps indicating a movement toward more global and connectivity driven computations required for later learning.

While there exist these relative and dynamic differences between activity and connectivity, the reorganization of activity and connectivity are both increasingly important to learning performance as learning progresses, as exhibited by higher learning coefficients for activity ($t=6.81$, $-\log10(p)=10$, $df=412$) and connectivity ($t=7.74$, $-\log10(p)=13$, $df=412$) on day four compared to day one. Interestingly, connectivity learning coefficients ($t=-3.16$, $p=0.003$, $df=48$), but not activity learning coefficients ($t=-1.75$, $p=0.087$, $df=48$), at value-associated regions increase significantly during learning from day one to day four(Figure 7d,e). Moreover, in general, connectivity reorganizations tracked value learning performance more than activity reorganizations, as indicated by higher connectivity than activity learning coefficients during learning ($t=2.41$, $p=0.016$, $df=1654$). Finally, we observed a very weak correlation between each region's learning coefficients for activity and connectivity ($r = 0.07$, $p = 0.001$, $df = 1650$), demonstrating their independence. These results further suggest that, subject's performance is enhanced when activity and connectivity reorganize to a group-common pattern, particularly for connectivity at regions associated with value and during later learning.

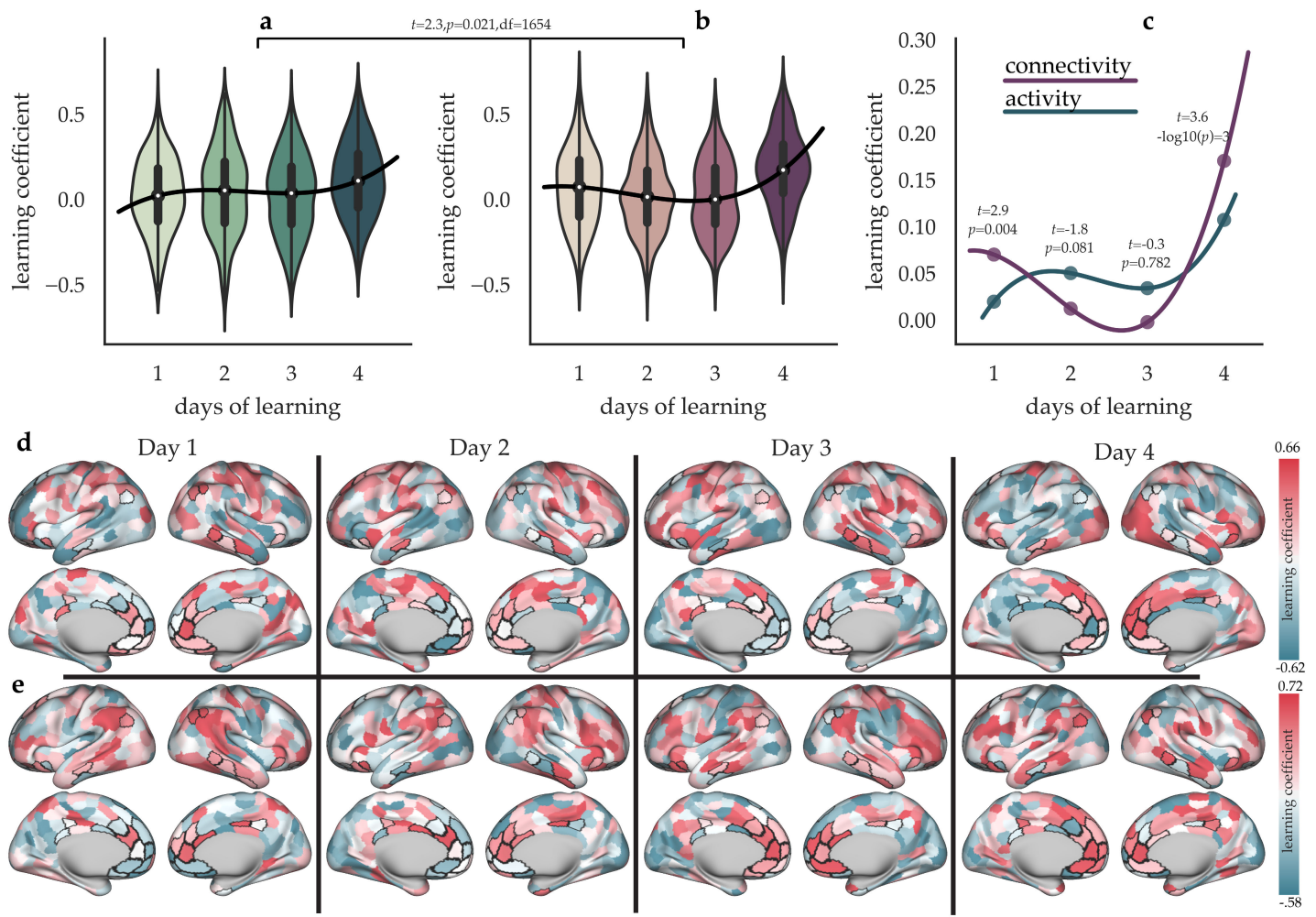

Figure 7 | **Activity and connectivity similarity explains subject learning performance on different days**. **a**, For each day and for each brain region, we calculated the Pearson correlation coefficient, $r$, between that region's activity reorganization across subjects and task performance across subjects. The distribution for these values, which we call learning coefficients, is shown for each day. **b**, As in panel **a**, but for connectivity reorganizations. **c**, The curves of learning coefficients for activity and connectivity across subjects for each day. **d**, The values from panel **a** shown on the cortical surface. **e**, The values from panel **b** shown on the cortical surface. As learning progresses, the relationship between activity and connectivity reorganizations and task performance strengthens, particularly for connectivity reorganizations and performance at value regions.

Finally, we reasoned that activity learning coefficients should be randomly distributed around zero during the resting-state, while patterns of connectivity critical to learning and formed during learning could persist into a resting state that occurs after training. Indeed, activity learning coefficients during the resting-state were indistinguishable from zero (one sample $t$=-1.2, $p$=0.21, df=1655). Further, learning coefficients were higher for connectivity than for activity during each day of resting-state, and significantly so on days two and four ($t$=9.1,14.2, -log10($p$)>10, df=412). Interestingly, the temporal aspect of connectivity learning coefficients was present in the resting-state; connectivity learning coefficients increased from days one to four ($t$=12.49, -log10($p$)=30, df=412), while activity learning coefficients decreased ($t$=3.02, $p$=0.003, df=412). Moreover, connectivity learning coefficients during resting-state were higher than connectivity learning coefficients during learning, ($t$=3.72, -log10($p$)=4, df=1654), while activity learning coefficients showed the opposite pattern ($t$=6.97, -log10($p$)=11, df=1654). Finally, each region's mean connectivity learning coefficients were correlated between learning and resting-state state; this was not true for activity learning coefficients (connectivity, $r$=0.25, -log10($p$)=7, df=412; activity, $r$=0.04, $p$=0.33, df=412). In sum, these results demonstrate that, as expected, connectivity, but not activity, during the resting-state, relates to learning performance. Moreover, learning induces a reorganization of resting-state connectivity, but not resting state activity. In fact, as the connectivity learning coefficients were higher during the resting-state, the specific patterns of connectivity reorganization that persist from learning into the resting-state are most relevant to learning performance.

**Discussion**

We sought to understand how activity and connectivity reorganize to a common functional architecture during learning by measuring the extent to which learning induces similarities in activity and connectivity across subjects, and by comparing our observations obtained during task to those obtained during the resting state. While connectivity reorganizes more than activity at brain regions previously associated with value learning, in general, the same set of brain regions exhibits reorganization in both activity and connectivity. However, activity reorganization peaks locally in the brain and early in learning, while connectivity reorganization slowly increases throughout learning and is more spatially distributed across the brain. Similarly, early activity reorganizations, and later connectivity reorganizations, are critical to learning performance. Finally, functional connectivity and activity reorganize the most in primary sensory communities, with functional connectivity additionally reorganizing in the fronto-parietal and default mode communities. These results are precisely in line with the modular theory of whole brain network computation, in which the function of somatomotor regions is largely explained by local activity patterns; in contrast, associative regions, primarily located in fronto-parietal cortex, execute integrative processing by modulating whole brain connectivity[2,27–33]. More generally, the local aspect of activity reorganizations and the spatially distributed aspect of connectivity reorganizations are in line with the idea that brain activity reflects local processing, while brain connectivity reflects information transfer between distant regions[29,30].

**Temporal Dynamics of Activity-Connectivity Reorganization**. While individuals can display a global reorganization of activity but not connectivity, *and vice versa*, both are increasingly important for performance as learning progresses. The extent to which reorganization in activity or connectivity contributes to performance follows a specific temporal pattern, in which connectivity reorganization tracks performance on the first day, then activity reorganization tracks performance on days two and three, and, finally, connectivity reorganization tracks performance on the final day of learning. Interestingly, this is the same temporal pattern of reorganizations in general, where activity reorganizations peak early in learning, while connectivity reorganizations slowly increase throughout learning. In general, these temporal patterns of increased connectivity reorganization to a common functional architecture trace out a trajectory reminiscent of a traditional learning curve where performance increases steeply at first before plateauing later[34]. It is interesting to speculate that changes in activity provide support for the early steep increases in performance, and changes in connectivity provide support for the later fine-tuning of behavior when performance gains from trial to trial are relatively small[35–37]. The early role of activity in explaining performance could in part be due to the fact that early stimulus encoding produces activity patterns that represent the objects being learned, and the accuracy and differentiation between those representations is critical for accurate responses in the task[38,39].

**Spatial Dynamics of Activity-Connectivity Reorganization.** Value learning induces more spatially distributed reorganization in activity early in learning, and in connectivity later in learning. This spatial pattern is remarkably similar to the temporal patterns described in the previous section. Intuitively, as learning progresses, the regions that are reorganizing connectivity become less spatially clustered together and more spatially distributed across the brain. Critically, we did not observe this spatiotemporal dynamic during the resting state, demonstrating its task relevance. Our findings are consistent with the notion that connectivity, at the regional macro-scale studied here, is more relevant to global information processing between spatially distant regions, while activity is likely more involved in spatially circumspect information processing. Also in line with this result, we observed a cascade of connectivity reorganization that peaked on days two, three, and four, of learning across three distinct sets of regions. This cascade effect was not present for activity reorganization.

It has previously been demonstrated that the ventral temporal cortex is critical to visual categorization of objects[40], functional connectivity between fronto-temporal regions to fronto-parietal is critical to value learning[19], and functional connectivity in the temporal cortex in general is critical for memory consolidation[23,24,41]. Our results here support the notion that functional connectivity is critical to memory consolidation[23,24,41], adding that consolidation, here, of the value of particular objects, might occur via the spread of reorganization to a common functional architecture that is induced by what the subjects are learning. More speculatively, it is possible that the temporal cortex regions of the default mode system, possibly coordinated by or driven by reorganization in the fronto-parietal community, reorganize late in learning specifically to support the consolidation of object values. While these regions are likely specific to value learning reorganization, there might be a more general phenomenon across the brain and tasks, in which tasks induce a more widespread reorganization in functional connectivity than activity.

**Activity-Connectivity Reorganization in the Resting State**. It is important to note that the dynamics of activity and connectivity that we observed during the task were strikingly different from the dynamics that we observed during rest. Interestingly, only connectivity reorganization during rest tracked learning performance, suggesting that "sticky" connectivity – that persisting from a task into the following rest period – contains information about individual differences in learning. Others have observed that, while task-dependent functional connectivity modes cluster around the connectivity architecture observed during rest[42], tasks can induce critical changes that persist in resting-state connectivity. For example, amygdala–hippocampal connectivity at rest after fear learning predicts later fear responses[43], the persistence of hippocampal multivoxel patterns and functional connectivity into post-encoding rest is related to memory performance[44,45], and perceptual training reconfigures post-task resting-state functional connectivity[46]. Interestingly, in some cases, the effects of cognitive effort on neural dynamics carry over into the resting state for up to 20 minutes[47]. Our findings add to this literature by demonstrating that individual differences in the degree to which the task has imprinted on the following resting state track individual differences in value learning.

**Future Work.** While our study addressed a single type of learning, the principles that we observed need not necessarily be constrained to value learning but may instead be generally relevant to related learning processes. It would be interesting in future work to test the hypothesis that other types of learning involve early activity reconfigurations as performance changes swiftly, with later connectivity reconfigurations as performance changes slowly. Such an investigation would require prolonged imaging throughout an extended training regimen. It would further be interesting to determine whether this phenomenon occurs at other spatial scales of investigation, such as in single lobes, cognitive systems, or regions. If so, it would join other features–such as brain modularity– that are scale free in nature[48]. By extending our investigation in this way, we would be able to better determine the generalizability of our observations, both across learning scenarios and across scales of brain physiology.

**Materials and Methods**

**Experimental setup and procedure.** Participants learned the monetary value of 12 novel visual stimuli in a reinforcement learning paradigm. Learning occurred over the course of four MRI scan sessions, each session occurring on one of four consecutive days. The novel stimuli were 3-dimensional shapes generated with a custom built MATLAB toolbox. Publicly available code is located here: http://github.com/saarela/ShapeToolbox. ShapeToolbox allows the generation of three-dimensional radial frequency patterns by modulating basis shapes, such as spheres, with an arbitrary combination of sinusoidal modulations in different frequencies, phases, amplitudes, and orientations. A large number of shapes were generated by selecting combinations of parameters at random. From this set, we selected twelve that were considered to be sufficiently distinct from one another. A different monetary value, varying from $1.00 to $12.00 in integer steps, was assigned to each shape. These values were not correlated with any parameter of the sinusoidal modulations, so that visual features were not informative of value.

Participants completed approximately 20 minutes of the main task protocol on each scan session, learning the values of the 12 shapes through feedback. The sessions were comprised of three scans of 6.6 minutes each, starting with 16.5 seconds of a blank gray screen, followed by 132 experimental trials (2.75 seconds each), and ending with another period of 16.5 seconds of a blank gray screen. Stimuli were back-projected onto a screen viewed by the participant through a mirror mounted on the head coil and subtended 4 degrees of visual angle, with 10 degrees separating the center of the two shapes. Each presentation lasted 2.5 seconds and, at any point within a trial, participants entered their responses on a 4-button response pad indicating their shape selection with a leftmost or rightmost button press. Stimuli were presented in a pseudorandom sequence with every pair of shapes presented once per scan.

Feedback was provided as soon as a response was entered and lasted until the end of the stimulus presentation period. Participants were randomly assigned to two groups depending on the type of feedback received. In the RELATIVE feedback case, the selected shape was highlighted with a green or red square, indicating whether the selected shape was the most valuable of the pair or not, respectively. In the ABSOLUTE feedback case, the actual value of the selected shape (with variation) was displayed in white font. After each run, both groups received feedback about the total amount of money accumulated in the experiment up to that point. The experimental protocol has been reported previously[5,19].

**MRI data collection and preprocessing.** Magnetic resonance images were obtained at the Hospital of the University of Pennsylvania using a 3.0 T Siemens Trio MRI scanner equipped with a 32-channel head coil. T1-weighted structural images of the whole brain were acquired on the first scan session using a three-dimensional magnetization prepared rapid acquisition gradient echo pulse sequence with the following parameters: repetition time (TR) 1620 ms, echo time (TE) 3.09 ms, inversion time 950 ms, voxel size 1 mm by 1 mm by 1 mm, and matrix size 190 by 263 by 165. To correct geometric distortion caused by magnetic field inhomogeneity, we also acquired a field map at each scan session with the following parameters: TR 1200 ms, TE1 4.06 ms, TE2 6.52 ms, flip angle 60°, voxel size 3.4 mm by 3.4 mm by 4.0 mm, field of view 220 mm, and matrix size 64 by 64 by 52. In all experimental runs with a behavioral task, T2*-weighted images sensitive to blood oxygenation level-dependent contrasts were acquired using a slice accelerated multi-band echo planar pulse sequence with the following parameters: TR 2000 ms, TE 25 ms, flip angle 60°, voxel size 1.5 mm by 1.5 mm by 1.5 mm, field of view 192 mm, and matrix size 128 by 128 by 80. In all resting state runs, T2*-weighted images sensitive to blood oxygenation level-dependent contrasts were acquired using a slice accelerated multi-band echo planar pulse sequence with the following parameters: TR 500 ms, TE 30 ms, flip angle 30°, voxel size 3.0 mm by 3.0 mm by 3.0 mm, field of view 192 mm, matrix size 64 by 64 by 48.

Cortical reconstruction and volumetric segmentation of the structural data was performed with the Freesurfer image analysis suite[49]. Boundary-Based Registration between structural and mean functional image was performed with Freesurfer bbregister[50]. Preprocessing of the fMRI data was carried out using FEAT (FMRI Expert Analysis Tool) Version 6.00, part of FSL (FMRIB's Software Library, www.fmrib.ox.ac.uk The following pre-statistics processing was applied: EPI distortion correction using FUGUE, motion correction using MCFLIRT[51], slice-timing correction using Fourier-space time series phase-shifting, non-brain removal using

BET[52], grand-mean intensity normalization of the entire 4D dataset by a single multiplicative factor, and highpass temporal filtering via Gaussian-weighted least-squares straight line fitting with $\sigma = 50.0s$. Nuisance time series were voxelwise regressed from the preprocessed data. Nuisance regressors included (i) three translation (X, Y, Z) and three rotation (pitch, yaw, roll) time series derived by retrospective head motion correction (R = [X, Y, Z, pitch, yaw, roll]), together with the first derivative and square expansion terms, for a total of 24 motion regressors[53]); (ii) the first five principal components of non-neural sources of noise, estimated by averaging signals within white matter and cerebrospinal fluid masks, obtained with Freesurfer segmentation tools, and removed using the anatomical CompCor method (aCompCor)[54]; and (iii) a measure of a local source of noise, estimated by averaging signals derived from the white matter region located within a 15 mm radius from each voxel, using the ANATICOR method[55]. Global signal was not regressed out of voxel time series.[56,57]

We parcellated the brain into 400 cortical regions defined by a cortical parcellation[16] and 14 sub-cortical regions defined by the structural Harvard-Oxford atlas of the fMRIB[17,18]. We warped the MNI152 regions into subject-specific native space using FSL FNIRT and nearest-neighbor interpolation and calculated the average BOLD signal across all gray matter voxels within each region. The participant's gray matter voxels were defined using the anatomical segmentation provided by Freesurfer, projected into the subject's EPI space with bbregister. For each individual scan, we extracted regional mean BOLD time series by averaging voxel time series in each of the 414 regions of interest.

**Brain network construction.** To perform network analyses one must define the two most fundamental elements of the network: nodes and edges. These two elements are the building blocks of networks and their accurate definitions are very important for any network models[58]. The standard method of defining nodes in the field of network neuroscience is to consider neuroimaging data such as fMRI and apply a structural atlas or parcellation that separates the whole brain volume into different regions defined by known anatomical differences[59]. A network node thus represents the collection of voxels within a single anatomically defined region. A network edge reflects the statistical dependency between the activity time series of two nodes. Each region's activity is given by the mean time series across all voxels within that region.

The edge weights that link network nodes were given by the wavelet transform coherence (WTC)[60], smoothed over time and frequency to avoid bias toward unity coherence. We chose to use wavelets, as wavelet analysis is multiresolution in nature and is adaptable to nonstationary or local features in data, common in fMRI times series[61]. Moroever, we have shown that wavelet coherence FC estimates had substantially lower fraction of edges significantly correlated with motion compared to other methods[62]. Specifically, we use Morlet wavelets with coefficients given by:

$$w(t,f) = (\sigma_t \sqrt{(\pi)})^{-\frac{1}{2}} e^{-i 2\pi f t} e^{-\frac{t^2}{2\sigma_t^2}}, \quad (1)$$

where $f$ is the center frequency and $\sigma_t$ is the temporal standard deviation. The time-frequency estimate, X(t,f) of time series x(t) was computed by a convolution with the wavelet coefficients:

$$X(t,f) = x(t) * w(t,f). \quad (2)$$

We selected the central frequency of 1/12 Hz corresponding to a spectral width of 0.05 to 0.11 Hz for full width at half maximum. Then the wavelet transform coherence between two time series x(t) and y(t) is defined as follows:

$$TC^2(f,t) = \frac{|S(s^{-1} X_{xy}(t,f)|^2}{S(s^{-1}|X_x(t,f)|^2) \cdot S(s^{-1}|X_y(t,f)|^2)}, \quad (3)$$

where $X_{xy}$ is the cross-wavelet of $X_x$ and $X_y$, s is the scale which depends on the frequency[63,64], and S is the smoothing operator. This definition closely resembles that of a traditional coherence, with the marked difference that the wavelet coherence provides a localized correlation coefficient in both time and frequency. Higher scales are required for lower frequency signals and in this study, we used s=32 for the smoothing operation. This procedure was repeated for all pair of regions yielding the 414 by 414 adjacency matrix, A, representing the functional connectivity between brain regions.

**Network modularity.** In network neuroscience, the term modularity can be used to refer to the concept that brain regions cluster into modules or communities[65]. These communities can be identified computationally using machine learning techniques in the form of community detection algorithms[66]. A community of nodes is a group of nodes that are tightly interconnected. In this study, we implemented a generalized Louvain-like community detection algorithm[25,67] that considers multiple adjacency matrices as slices of a multilayer network, and which then produces a partition of brain regions into modules that reflects each subject's community structure across the multiple stages of learning instantiated in the four days of task practice. The multilayer network was constructed by connecting the adjacency matrices of all scans and subjects with interlayer links. We then maximized a multilayer modularity quality function, $Q$, thereby obtaining a partition of nodes into communities in a way that maximizes the strength of intra-community connections[67]:

$$Q = \frac{1}{2\mu} \sum_{ijs}[(A_{ijs} - \gamma_s V_{ijs})\delta_{sr} + \delta_{ij}\omega_{jsr}]\delta(g_{is}, g_{jr}) \tag{4}$$

where $A_{ijs}$ is the ij$^{th}$ element of the adjacency matrix of slice s, and element $V_{ijs}$ is the component of the null model matrix tuned by the structural resolution parameter $\gamma$. In this study, we set $\gamma=1$, which is the standard practice in the field when no a priori hypotheses exist to otherwise inform the choice of $\gamma$. We employed the Newman-Girvan null model within each layer by using $V_{ijs} = \frac{k_{is}k_{js}}{2m_s}$, where $k$ is the total edge weight and $m_s$ is the total edge weight in slice s. The interslice coupling parameter, $\omega_{jsr}$, is the connection strength of the interlayer link between node j in slice s and node j in slice r, and the total edge weight in the network is $\mu = \frac{1}{2}\sum_{jr} \kappa_{jr}$. The node strength, $\kappa_{jr}$, is the sum of the intraslice strength and interslice strength: $\kappa_{jr} = k_{jr} + c_{jr}$, and $c_{jr} = \sum_s \omega_{jrs}$. In this study, we set $\omega = 1$, which is the standard practice in the field when no a priori hypotheses exist to otherwise inform the choice of $\omega$. Finally, the indicator $\delta(g_i, g_j) = 1$ if nodes i and j are assigned to the same community, and is 0 otherwise.

We obtained a partition of the brain into communities for each scan and subject. To obtain a single representative partition of brain regions into distinct communities, we computed a module allegiance matrix[4,22,68], whose ij$^{th}$ entry represents the probability that region i and region j belong to the same community across scans and participants. We then applied single-layer modularity maximization to this module allegiance matrix to obtain a single partition of brain regions into consensus modules.

**Statistical Methods: Inter-subject correlation and inter-subject functional connectivity.** We used the Pearson correlation coefficient to compute the inter-subject correlation (inter-subject correlation) of each brain region for each subject. First, we calculated the region-wise temporal correlation between every pair of subjects as:

$$r_{ij} = \frac{\sum_{t=1}^{N}[(x_i(t) - \bar{x}_i)(x_j(t) - \bar{x}_j)]}{\sqrt{\sum_{t=1}^{N}(x_i(t) - \bar{x}_i)^2 \sum_{t=1}^{N}(x_j(t) - \bar{x}_j)^2}} \tag{6}$$

where $N$ is the number of points in the time series data, and where $r_{ij}$ is the correlation coefficient of a region between the times series $x_i$ and $x_j$ of the $i$th and $j$th subjects, respectively.

To test the statistical significance of the correlation between the fMRI BOLD signals of a single region from two subjects, we performed a fully non-parametric permutation test with 10,000 randomizations. This test accounts for slow-scale autocorrelation structure in the BOLD time series by removing phase information from each BOLD signal through Fourier phase randomization. We repeated this procedure 10,000 times to obtain a null distribution of the maximum noise correlation values, and we defined the threshold for a pair correlation as the $q \times 100^{th}$ percentile of the null distribution of maximum values.

To obtain the inter-subject correlation, we averaged only significant correlation values out of 414 correlation values, $r_{ij}$ from all subject pairs, to obtain one inter-subject correlation for each region:

$$ISC = \frac{2}{M(M-1)} \sum_{i=1}^{M} \sum_{j=2, j>1}^{M} r_{ij}, \tag{7}$$

where $M$ is the number of subjects. This same procedure was followed for each scan, for each day, and for both rest and task conditions.

Inter-subject functional connectivity (inter-subject functional connectivity) was obtained from the functional connectivity (FC) matrices of all of the subjects, and can be thought of as an estimate of the correlation in FC between one subject and all other subjects. Specifically, we computed the inter-subject functional connectivity of each subject as the correlation between the single subject FC matrix and the average of all other subject-specific FC matrices as

$$ISFC_i = \frac{1}{N} A_i [\frac{1}{n-1} \sum_{j \neq i} A_j], \tag{8}$$

where A is the functional connectivity matrix of a subject, and where n and N are the total number of subjects and total number of regions, respectively.

**Statistical Methods: Moran's I.** Moran's *I* is a measure of spatial autocorrelation. For example, if values 0 and 1 are perfectly dispersed in a checkerboard pattern, Moran's I is -1. If all 0's are stacked on one side and all 1's are stacked on the other, Moran's I is 1. A Moran's I of 0 is random. Here, we take the inverse of Moran's I as a measure of how dispersed values of ISFC and ISC are across the cortex, an inverse measure of how spatially clustered the values are. Moran's I is calculated as follows:

$$I = \frac{N}{W} \frac{\Sigma_i \Sigma_j w_{ij}(x_i - \bar{x})(x_j - \bar{x})}{\Sigma_i (x_i - \bar{x})^2},$$

where $N$ is the number of spatial units indexed by $i$ and $j$; $x$ is the variable of interest; here, ISFC or ISC, $\bar{x}$ is the mean of x; $w_{ij}$ is a matrix of spatial weights with zeroes on the diagonal. These spatial weights are Euclidian distance from the center of parcel to every other parcel. $W$ is the sum of all $w_{ij}$.

**Statistical Methods: Assumptions of normality.** Here, we tested a variety of group differences using the student's *t*-test, which assumes a normal distribution. Both ISFC and ISC approximate a normal distribution by visual inspection (Supplemental Figure 1).[67]

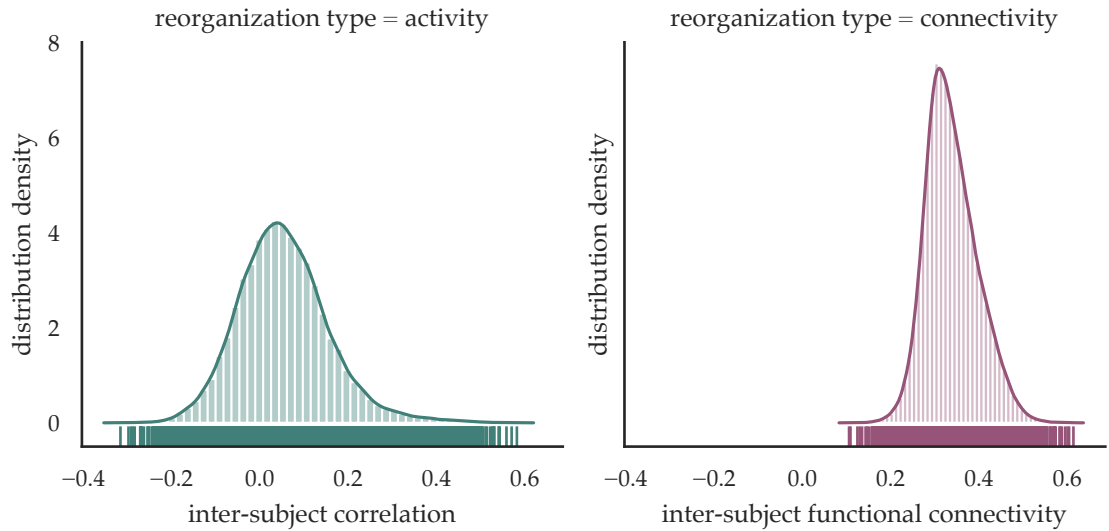

**Supplemental Figure 1 | Distributions of ISC and ISFC.** Univariate distribution of ISC and ISFC are shown, with a rugplot underneath showing individual datapoints; this plot combines a traditional histogram of the values with a kernel density estimation function. While ISC is mostly normally distributed, it has a slight kurtosis for upper values. In contrast, while ISFC is mostly normally distributed, it has a slight skew to the lower values. Despite these slight deviations from normality, the data approximate normal distributions.

**Methodological Considerations.** The number of subjects analyzed here is smaller than large-scale data collections[69]. However, given the four-day experimental design involving the same subjects, totaling 80 MRI scans, this dataset nevertheless represents a rich opportunity to analyze the temporal evolution of reorganization of brain connectivity and activity during learning. Moreover, ideally, learning could be sampled in the scanner more than four times. However, we believe that, given 80 scans, we achieved an optimal balance between sampling learning enough in the temporally, but also sampling enough subjects. Recent research has shown the utility in analyzing "deeply phenotyped" subjects[6,70,71]. We hope that future large-scale data collections will include similar experiments. Finally, we chose a simple and intuitive method to measure the similarity of brain activity and connectivity across subjects, but certainly more sophisticated methods could be developed and will likely uncover results that are complimentary to the results reported here.

**Acknowledgements.** We thank David Lydon-Staley for helpful feedback on an earlier version of this manuscript. D.S.B. acknowledges support from the John D. and Catherine T. MacArthur Foundation, the Alfred P. Sloan Foundation, the ISI Foundation, the Paul G. Allen Foundation, the Army Research Laboratory (W911NF-10-2-0022), the Army Research Office (Bassett-W911NF-14-1-0679, Grafton-W911NF16-1-0474, DCIST- W911NF-17-2-0181), the Office of Naval Research, the National Institute of Mental Health (2-R01-DC009209-11, R0-MH112847, R01-MH107235, R21-MH106799, R01-MH113550), the National Institute of Child Health and Human Development (1R01-HD086888-01), National Institute of Neurological Disorders and Stroke (R01-NS099348), and the National Science Foundation (BCS-1441502, BCS-1430087, NSF PHY-1554488 and BCS-1631550). The content is solely the responsibility of the authors and does not necessarily represent the official views of any of the funding agencies.

**Citation Diversity Statement** Recent work in neuroscience and other fields has identified a bias in citation practices such that papers from women and other minorities are under-cited relative to the number of such papers in the field[72–74]. Here we sought to proactively consider choosing references that reflect the diversity of the field in thought, form of contribution, gender, race, geographic location, and other factors. We used open source code[75] an automatic classification of gender (gender-api.com) based on the first names of the first and last authors, with possible combinations including man-man, man-woman, woman-man, woman-woman. Excluding self-citations to the first and senior authors of our current paper, the references contain 56% (n=32) man-man, 12.5% (n=7) man-woman, 17.5% (n=10) woman-man, and 14% (n=8) woman-woman. Expected proportions reported in

Dworkin et al. (2020)[72] for 5 high-impact neuroscience journals are, respectively: 58.4%, 9.4%, 25.5%, and 6.7%. We look forward to future work that could help us to better understand how to support equitable practices in science.